\documentclass[twocolumn,showpacs,superscriptaddress]{revtex4}% Physical Review 
\usepackage{graphicx}% Include figure files
\usepackage{dcolumn}% Align table columns on decimal point
\usepackage{bm}% bold math

\begin{document}

\title{Magnetization of ultrathin (Ga,Mn)As layers}

\author{R. Mathieu\cite{newaddress}}
\affiliation{Department of Materials Science, Uppsala University, Box 534, SE-751 21 Uppsala, Sweden}

\author{B. S. S{\o}rensen}
\affiliation{Niels Bohr Institute fAFG, {\O}rsted Laboratory, University of Copenhagen, DK-2100 Copenhagen, Denmark}

\author{J. Sadowski}
\affiliation{Niels Bohr Institute fAFG, {\O}rsted Laboratory, University of Copenhagen, DK-2100 Copenhagen, Denmark}
\affiliation{MAX-Lab, Lund University, SE-221 00 Lund, Sweden}
\affiliation{Institute of Physics, Polish Academy of Sciences, Al. Lotnikow 32/46, PL-02668 Warszawa, Poland}

\author{U. S\"odervall}
\affiliation{Chalmers University of Technology, SE-412 96 G\"oteborg, Sweden}

\author{J. Kanski}
\affiliation{Chalmers University of Technology, SE-412 96 G\"oteborg, Sweden}

\author{P. Svedlindh}
\affiliation{Department of Materials Science, Uppsala University, Box 534, SE-751 21 Uppsala, Sweden}

\author{P. E. Lindelof}
\affiliation{Niels Bohr Institute fAFG, {\O}rsted Laboratory, University of Copenhagen, DK-2100 Copenhagen, Denmark}

\author{D. Hrabovsky}
\affiliation{Laboratoire de Physique de la Mati\`ere Condens\'ee, UMR CNRS 5830, INSA, 31077 Toulouse, France}

\author{E. Vanelle}
\affiliation{Laboratoire de Physique de la Mati\`ere Condens\'ee, UMR CNRS 5830, INSA, 31077 Toulouse, France}

\date{\today}

\begin{abstract}
Kerr rotation and Superconducting QUantum Interference Device (SQUID) magnetometry measurements were performed on ultrathin (Ga$_{0.95}$Mn$_{0.05}$)As layers. The thinner layers (below 250 \AA) exhibit magnetic properties different than those of thicker ones, associated with different microstructure, and some degree of inhomogeneity. The temperature dependence of the field-cooled-magnetization of the layers is recorded after successive low temperature annealings. While the Curie temperature of the thicker layer (250 \AA) is nearly unchanged, the critical temperature of the thinner layers is enhanced by more than 23 K after two annealings. Secondary Ion Mass Spectrometry (SIMS) experiments on similar layers show that Mn is displaced upon annealing. The results are discussed considering a possible segregation of substitutional and interstitial Mn atoms at the surface of the (Ga,Mn)As layers.
\end{abstract}

\pacs{75.50.Pp, 75.70.Ak}

\maketitle

\section{Introduction}

The discovery of ferromagnetism in (Ga,Mn)As is rather recent\cite{ohno1} and the exact mechanisms of the magnetic interaction are still under discussion. The appearance of ferromagnetism in (Ga,Mn)As is usually related to the magnetic exchange interaction between charge carriers (holes) and localized magnetic moments\cite{dietl} of the Mn$_{\rm Ga}$ atoms. The most popular theories include hole mediated ordering of the local Mn spins via a Ruderman-Kittel-Kasuya-Yosida (RKKY) interaction\cite{dietl}, or the competition between indirect exchange mechanisms such as double and super exchange\cite{akai}.

Due to the limited solubility of Mn in bulk GaAs, (Ga,Mn)As is grown as thin layers, by means of low temperature molecular beam epitaxy\cite{ohno1b,jan1} (LTMBE). It is observed that the total magnetic moment, the Curie temperature ($T_c$) associated to the paramagnetic-to-ferromagnetic transition, and the electrical properties of such (Ga,Mn)As layers are crucially dependent on the amount of Mn\cite{ohno2} and its distribution over the different positions in the crystal lattice of the GaAs host\cite{yu,blino}, as well as on the concentration of other defects compensating Mn acceptors\cite{munekata,masek}. The hole concentration of the layers is only a fraction of the expected one, and the degradation of the magnetic and electrical properties is often related to presence of compensating donor defects, such as As or Mn interstitials, Ga vacancies, or As$_{\rm Ga}$ antisites\cite{grandi,pavel} in the structure.

Ohno et al. reported\cite{ohno2} a $T_c$ of 110 K for 5.3 \% Mn doped (Ga,Mn)As. This value of 110 K could be refered to as the ``110 K limit'', as many research groups - until recently, as we will discuss in the following - could reach this value, without overcoming it. In order to increase $T_c$ toward room temperature, it is necessary to increase the number of free charge carriers mediating the ferromagnetic interaction. It is difficult to increase the Mn concentration above 10 \% \cite{jan2}. One can instead improve the magnetic properties of (Ga,Mn)As by reducing the amount of compensating defects. For example, it has been predicted theoretically that As$_{\rm Ga}$ antisites could be transformed in [As intersticial - Ga vacancy] pair upon illumination\cite{sanvito}, reducing the hole compensation. It was also demonstrated that post-growth annealing could displace interstitial Mn atoms, breaking the passivating [Mn interstitial - Mn$_{\rm Ga}$] pairs\cite{yu}.

Several groups have indeed been successful in increasing $T_c$ by annealing the layers after growth. Hayashi et al. \cite{hayashi} could increase the $T_c$ of their 2000 {\AA} thick (Ga$_{0.95}$Mn$_{0.05}$)As layer from 45 K to 95 K by annealing it 50$^\circ$C above its growth temperature. Later, Potashnik et al.\cite{potashnik} obtained a $T_c$ of 110 K in a 1100 {\AA} thick (Ga$_{0.92}$Mn$_{0.08}$)As layer shortly annealed at a temperature just above the growth temperature. They also observed that, above 5 \% of Mn, the $T_c$ of optimally annealed layers became independent of the Mn content\cite{potashnik2}, saturating at 110 K. Edmonds et al.\cite{gallagher} reported similar results in thinner layers, as well as an increase of the hole concentration by a factor of 5 in a (Ga$_{0.92}$Mn$_{0.08}$)As layer upon annealing. It has also been found recently\cite{brian,brianapl,ku,newedm} that the post growth annealing of (Ga,Mn)As is the most effective in a very specific thickness range. 

In the present article, we report low temperature annealing studies on thin layers of (Ga$_{0.95}$Mn$_{0.05}$)As, with thicknesses ranging from 250 {\AA} to 50 {\AA}. The effects of annealing depend on the (Ga,Mn)As layer thickness, inducing the largest changes in the magnetization curves and Curie temperatures of the thinnest layers. $T_c$ gradually increases after each annealing and the highest Curie temperature of 107 K is reached for the 150 {\AA} thick layer.  To our knowledge, we present the first direct magnetization measurements for (Ga,Mn)As layers with thicknesses in the range of 70  - 150 {\AA}. We hope that similar measurements appear in the literature, helping to disclose the microscopic origin of the unusual shape of the magnetization curves of such thin layers.

\section{Experimental}

Six (Ga$_{0.95}$Mn$_{0.05}$)As layers with thicknesses of 50, 70, 100, 150, 200, and 250 {\AA} were epitaxially grown by LTMBE on epi-ready GaAs(100) wafers\cite{brian}, and covered by a 30 {\AA} thick low temperature GaAs layer. This capping layer is necessary to prevent oxidation of the (Ga,Mn)As surface when the samples are taken out of the vacuum system. All layers were prepared under the same conditions, at a growth temperature of 230 $^\circ$C. The Kerr rotation was measured as a function of temperature, employing a He-Ne laser beam as a probe. Details on the experimental setup can be found in Ref. \onlinecite{kerr}. The field-cooled (FC) magnetization of the layers was recorded using a Quantum Design MPMS5 Superconducting QUantum Interference Device (SQUID) in a small in-plane applied magnetic field ($H$ = 20 or 50 Oe). The temperature dependence of the magnetization $M$($T$) of the ``as grown'' layers, as well as after 1, 2, and even 3 post-growth annealings was collected. Each post-growth annealing was performed at 240$^\circ$C for one hour in a nitrogen atmosphere\cite{note}. Secondary Ion Mass Spectrometry (SIMS) experiments were performed using a CAMECA IMS-6F instrument on a 1.2 $\mu$m (Ga$_{0.93}$Mn$_{0.07}$)As layer. The (Ga,Mn)As was covered with a 100 {\AA} thick low temperature GaAs capping layer. After growth, the sample was divided into separate pieces, which were annealed in high vacuum at 280$^\circ$C for 1 and 4 hours respectively. The two pieces were  analyzed in depth profiling mode after and before annealing. The primary ion beam was created by accelerating positive oxygen ions, O$_2^+$, with a voltage of 6.5 kV, corresponding to a net impact energy of 2 keV. The diameter of the ion beam was 50 $\mu$m and it was scanned over an area of 250 $\times$ 250 $\mu$m$^2$. The current intensity was circa 25 nA giving a sputter rate of $\sim$ 0.1 nm/s. The positive secondary ions of $^{55}$Mn,$^{71}$Ga and $^{75}$As were analyzed. Only ions from the central area of sputtering with a diameter of 50 $\mu$m were analyzed. Molecular mass discrimination was applied by using an energy offset of 100 V.

\section{Results and discussion}

\begin{figure}
\includegraphics[scale=0.51]{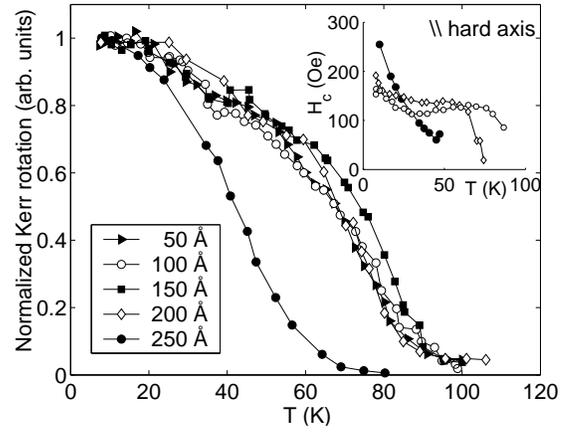}
\caption{Temperature dependence of the Kerr rotation for all layers. The temperature dependence of the coercivity of the layers with  250, 200 and 100 {\AA} is shown in inset.}
\label{fig0}
\end{figure}

\begin{figure}
\includegraphics[scale=0.52]{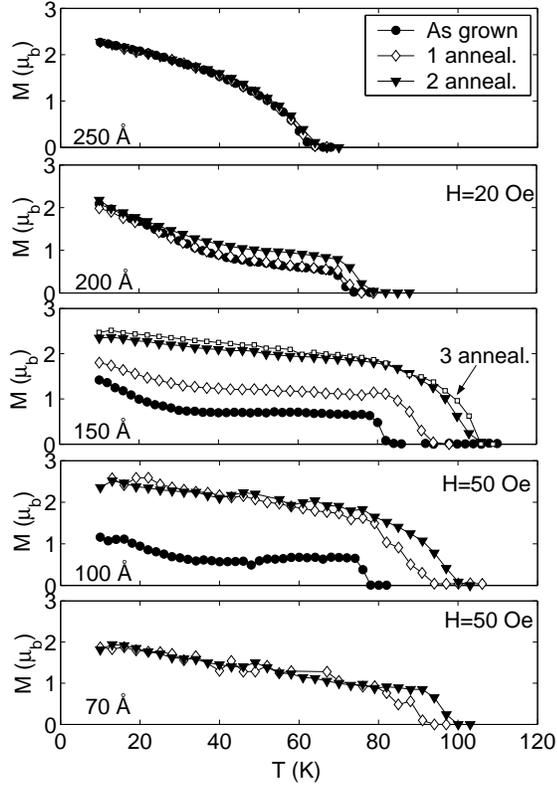}
\caption{Temperature dependence of the FC magnetization for all layers; as grown (filled circles), annealed once (open diamonds) and annealed twice (filled triangles). The magnetization is recorded on heating in a small magnetic field ($H$=20 Oe for 250, 200 and 150 {\AA}; $H$=50 Oe for 100 and 70 {\AA}). In the case of the 150 {\AA} thin layer, the FC magnetization recorded after a third annealing are included (open squares).}
\label{fig1}
\end{figure}

The magnetization of the thin layers was first investigated using the magneto-optical Kerr-effect. The main frame of Fig.~\ref{fig0} shows the temperature dependence of the Kerr rotation for the layers with 50, 100, 150, 200, and 250 {\AA}. As seen on the figure, magnetic ordering occurs at higher temperatures for the thinner layers. The temperature onset of ordering is similar for the thin layers, but this ordering is more pronounced for the 150 {\AA} thin layer. For the thicker (250 {\AA}) layer, the coercivity decreases rapidly with increasing temperature (see inset), while it remains constant over a relatively large temperature range in the case of the thinner layers. This indicates some magnetic inhomogeneity in the thinner samples. The random distribution of the Mn acceptors and the donor defects of (Ga,Mn)As affect the magnetic interaction locally, depending on the local hole-electron compensation (see below for more details). The shape of the magnetization curves of the layers is studied in more details using SQUID magnetometry. Figure~\ref{fig1} shows the temperature dependence of the FC magnetization for the different layer thicknesses. The magnetization curve of the as-grown 250 {\AA} thick layer is typical for ferromagnetic (Ga,Mn)As samples\cite{jan2}, with a $T_c$ of 63 K; $T_c$ is here determined by the onset of magnetic ordering in the $M$($T$) curves. In the case of thinner layers, the magnetization curves are broader, which again indicates some degree of magnetic inhomogeneity\cite{jan2}. The shape of the $M$($T$) curves is very similar to that predicted by Berciu and Bhatt in case of segregation of the Mn in ``Mn-rich'' and ``Mn-poor'' regions\cite{berciu}. In that case, the regions with larger Mn concentration have a larger density of holes, which in turn locally enhances the hole-spin interaction. $T_c$ is larger for the 150 {\AA} thick layer. The onset of ferromagnetism increases up to 83 K for this layer, but decreases for lower thicknesses. The magnetization of the as-grown 50 and 70 {\AA} thick layers was too small to be detected on the SQUID.

Successive low temperature annealings are performed on the layers. The $T_c$ of the thicker layers (250 and 200 {\AA}) is not significantly affected by a first annealing, showing an increase of only 0.4 and 2.3 K respectively. For the thinner layers, a large increase of $T_c$ is observed, and the $T_c$ of the 150 and 100 {\AA} thick layers increases by 11.3 and 15.7 K respectively. A significant magnetic signal appeared after heat treatment in the 70 {\AA} thick layer, and a $T_c$ of 92 K could be observed\cite{note2}. The values of $T_c$ obtained after this first annealing coincide with the values previously obtained by Hall measurements\cite{brian} on the same samples (also annealed once). The effects of a second annealing are similar, and further increases in $T_c$ are observed: The $T_c$ of the thicker layers increases by 1.3 and 3.6 K respectively, while a larger increase is obtained for the thinner layers. For (Ga,Mn)As thicknesses of 150, 100 and 70 {\AA}, an onset of ferromagnetism appears in the $M$($T$) curves at 107, 101, and 99.7 K respectively.

\begin{figure}
\includegraphics[scale=0.5]{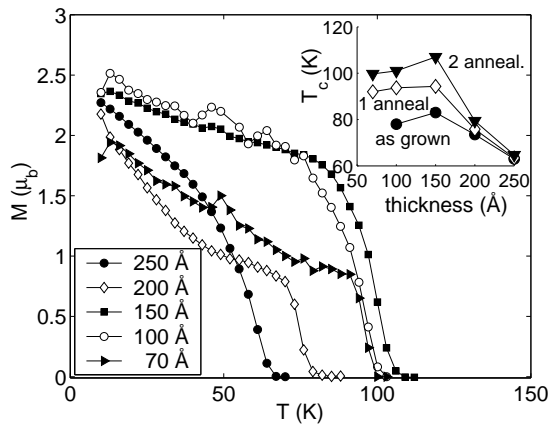}
\caption{Temperature dependence of the FC magnetization for all layers after two annealings; $H$=20 Oe for 250, 200 and 150 {\AA}; $H$=50 Oe for 100 and 70 {\AA}). The inset shows the dependence of the Curie temperature on the thickness of the layers.}
\label{fig2}
\end{figure}

These results are summarized in Fig.~\ref{fig2}, which shows the temperature dependence of the fied-cooled magnetization for all layers after two annealings. As seen in the insert, and as mentioned above, the effect of the annealings are larger for the smaller thicknesses. As seen in the main frame of Fig.~\ref{fig2}, the magnetization reaches a value of $\sim$ 2.5 $\mu_B$ at low temperatures for all layers. This indicates an increase of the total magnetic moment for all thicknesses, as well as possible changes in the coercivity of the layers, as observed by Kuryliszyn et al.\cite{kury}. In their study, the coercive field of $\sim$ 1000 - 3000 {\AA} thick (Ga,Mn)As layers was found to decrease by a factor of 2 after short annealings close to the growth temperature.

A third annealing is performed on the 150 {\AA} thick layer. As seen in Fig~\ref{fig1}, $T_c$ and the total magnetic moment are only marginally affected, indicating that the optimal annealing time for these layers is thus close to 3 hours. After three annealings, the shape of the magnetization curve of this layer was greatly affected, becoming similar to those obtained for thicker layers\cite{jan2}. Interestingly, we here reach the ``110 K limit'', which may be related to the largest amount of free holes\cite{yu,seong} accessible in such (Ga,Mn)As layers. This limit could be, as pointed out recently by Edmonds et al.\cite{newedm}, fixed by the details of the growth conditions, and the amount of defects already present before the post-growth annealings. It was recently demonstrated using the present set of samples that the number of carriers available, and thus the amount of defects, was the main factor determining $T_c$\cite{brianapl}. By performing longer annealings,  Edmonds et al.  could increase $T_c$ up to 140 K. Post-growth annealings allowed Ku et al. \cite{ku} to reach 150 K. They also observed that the larger $T_c$ were obtained for the thinner layers, with a maximum for a thickness of 150 {\AA}. 
 The different annealings rearrange the structure of the layers, displacing both the donor defects and the acceptor Mn ions. As$_{\rm Ga}$ antisites are easily formed during the growth of the (Ga,Mn)As layers. Sanvito et al.\cite{sanvito2} have shown that the magnetic interaction in (Ga,Mn)As was influenced by the amount of antisites, as well as by the position of this antisites with respect to the Mn ions. Mn interstitials are as well present in a non-negligible amount\cite{newgrandi}. Yu et al.\cite{yu} could relate the enhancement of the magnetic properties of (Ga,Mn)As upon annealing to the displacement of interstitial Mn atoms.

\begin{figure}
\includegraphics[scale=0.52]{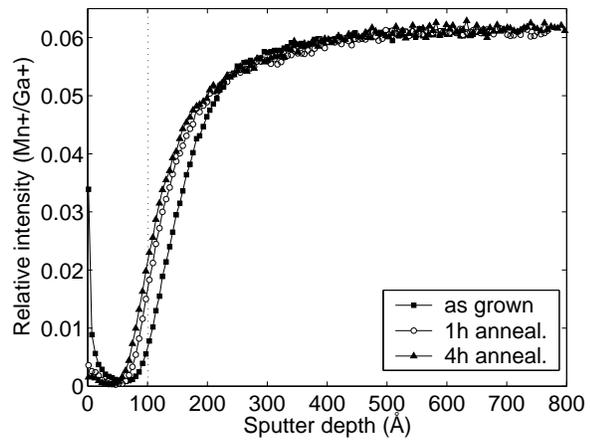}
\caption{SIMS Mn distribution profiles from the as-grown (Ga,Mn)As layer, and from its pieces annealed for 1 hour and 4  hours. The dotted line indicate the interface between the capping layer and (Ga,Mn)As.}
\label{fig4}
\end{figure}

Our present results may indicate a segregation of the Mn on the surface of the layers\cite{newcamp}.  Erwin and Petukhov\cite{erwin} have shown that the initial accommodation of Mn atoms should occur preferentially in interstitial sites, implying that such sites should be more abundant in the surface region. Experiments\cite{koeder} have indeed revealed a gradient in the concentration of holes across the thickness of (Ga,Mn)As layers. Although contrary to what one might expect from the just mentioned theoretical considerations, the density of holes is found to be highest in the surface region. These apparently contradicting results may be reconciled if one assumes that the Mn interstitials are mobile under the prevailing growth conditions, and that the surface region is gradually depleted of Mn interstitials by diffusion and surface segregation. Assuming further that the diffusion is slower than the typical growth rate, it is natural to expect that without post-growth annealing, the depletion process should be most efficient for relatively thin layers. Above a certain thickness, determined by the detailed growth/post-growth process, an equilibrium concentration profile of interstitials should be "frozen in". Such a scenario is consistent with our results, as well as the results obtained by Ku et al. \cite{ku}, showing that there is an optimum thickness of (Ga,Mn)As layers, above which the magnetic and transport properties degrade. It is also consistent with the observed efficiency of extremely long annealings demonstrated by Edmonds et al.\cite{gallagher} .
Although the various experimental results are strongly suggestive, the assumption of Mn diffusion must be verified experimentally. For this purpose we have carried out studies of the Mn concentration profile using Secondary Ion Mass Spectroscopy (SIMS).  The SIMS Mn distribution profiles from  an as-grown (Ga,Mn)As layer\cite{note3}, and from its pieces annealed for 1 hour and 4  hours are displayed in Fig.~\ref{fig4}. The data clearly confirm our assumption that Mn atoms are displaced in the annealing  process, not only within the (Ga,Mn)As layer, but also across the GaAs/(Ga,Mn)As interface. More detailed studies\cite{newjan} show that Mn diffuses to the capping layers,  probably in the form of Mn interstitials. Since these interstitials act as double donors\cite{blino,lars}, they tend to compensate the acceptor-type Mn atoms in substitutional sites. The unusual shape of the magnetization curves is related to the inhomogeneity of the thin layers. Varying concentration of Mn interstitials thus implies varying magnetic properties through the (Ga,Mn)As layer. 

\section{Conclusion}

The magnetic properties of thin layers of (Ga$_{0.95}$Mn$_{0.05}$)As are investigated using the Kerr effect and SQUID magnetometry. The effects of low temperature annealings are more pronounced for the thinner layers of (Ga,Mn)As. The results can be understood considering that the surface of (Ga,Mn)As layers host a larger number of interstitial Mn atoms, and is more sensitive to low temperature annealings than the volume of the layers. SIMS measurement do confirm the displacement of the Mn atoms during the annealing process.

While for thick layers, the latter dominates the magnetic response, for the thinner layers, the improvements of the surface upon annealing are unmasked. We believe that the optimization of the growth and post-growth (annealings) conditions of such thin layers could yield very large $T_c$ values. Since the present layers are sandwiched between two low temperature GaAs layers, having a lower density of defects, the magnetic properties of the (Ga,Mn)As layers could also be affected by interface effects. It is also important to note that such ultrathin layers certainly exhibit a complex magnetic domain structure\cite{fuku}, which is rearranged during the different heat treatments.

\begin{acknowledgments}

Financial support from the Swedish Research Council (VR) is acknowledged. This work was supported by the Danish Research Council of Engineering Sciences (STVF framework programme ``Nanomagnetism''), and the Danish Science Research Council (SNF framework programme ``Mesoscopic Physics''). The LTMBE growth of the samples was performed at MAX-Lab, Lund University. J. Sundqvist is acknowledged for his expert help.

\end{acknowledgments}

\newpage

\begin{thebibliography}{10}
\bibitem{note}
The optimum annealing conditions (annealing temperature and duration) will be investigated in a further study.
\bibitem{note2}
As in the Kerr effect measurements (see Fig.~\ref{fig0}), ferromagnetism could be detected in the 50 {\AA} (Ga,Mn)As layer in the Hall measurements[\onlinecite{brian}].
\bibitem{note3}
This sample was specially designed for SIMS measurements. Very thin (Ga,Mn)As samples are not suitable for such experiments. At low sputter depths the amplitude of SIMS signal has no direct connection to the measured ions concentration. The ratios of Mn to Ga signals are not affected and reflect correctly Mn to Ga concentration relations even at the very surface region.
\end{thebibliography}
\end{document}